\documentstyle[12pt,psfig]{article}
\begin{document}
\setlength{\headheight}{0in}
\setlength{\headsep}{0in}
\setlength{\topskip}{1ex}
\setlength{\textheight}{8.5in}
\setlength{\topmargin}{0.5cm}
\setlength{\baselineskip}{0.24in}
\catcode`@=11
\long\def\@caption#1[#2]#3{\par\addcontentsline{\csname
  ext@#1\endcsname}{#1}{\protect\numberline{\csname
  the#1\endcsname}{\ignorespaces #2}}\begingroup
    \small
    \@parboxrestore
    \@makecaption{\csname fnum@#1\endcsname}{\ignorespaces #3}\par
  \endgroup}
\catcode`@=12
\def\slashchar#1{\setbox0=\hbox{$#1$}           
   \dimen0=\wd0                                 
   \setbox1=\hbox{/} \dimen1=\wd1               
   \ifdim\dimen0>\dimen1                        
      \rlap{\hbox to \dimen0{\hfil/\hfil}}      
      #1                                        
   \else                                        
      \rlap{\hbox to \dimen1{\hfil$#1$\hfil}}   
      /                                         
   \fi}                                         %
\newcommand{\newc}{\newcommand}
\def\be{\begin{equation}}
\def\ee{\end{equation}}
\def\bea{\begin{eqnarray}}
\def\eea{\end{eqnarray}}
\def\simlt{\stackrel{<}{{}_\sim}}
\def\simgt{\stackrel{>}{{}_\sim}}
\begin{titlepage}
\begin{flushright}
{\setlength{\baselineskip}{0.18in}
{\normalsize
hep-ph/9705315 \\
SUSX-TH-97-008\\
UPR-0749-T\\
IEM-FT-156/97\\
May 1997\\
}}
\end{flushright}
\vskip 2cm
\begin{center}

{\Large\bf 
Cold Dark Matter Candidate in a Class of Supersymmetric Models
with an extra U(1)}

\vskip 1cm

{\large
B. de Carlos${}^{a,}$\footnote{Work supported by PPARC.} 
and J.R. Espinosa${}^{b,}$\footnote{Work supported by 
the Department of Energy, Grant DOE-EY-76-02-3071.} \\}

\vskip 0.5cm
{\setlength{\baselineskip}{0.18in}
{\normalsize\it ${}^a$ Centre for Theoretical Physics\\
           University of Sussex \\
	   Falmer, Brighton BN1 9QH, UK \\}
\vskip 4pt
{\normalsize\it ${}^b$Department of Physics and Astronomy  \\
	University of Pennsylvania \\
        Philadelphia PA 19104-6396, USA\\} }

\end{center}
\vskip .5cm
\begin{abstract}
In supersymmetric models whose gauge group includes an additional 
$U(1)$ factor at the TeV scale, broken by the VEV of an standard 
model singlet $S$, the parameter space can accommodate a very light 
neutralino not ruled out experimentally. This higgsino-like fermion, 
stable if $R$-parity is conserved, can make a good cold dark matter 
candidate. We examine the thermal relic density of this particle and
discuss the prospects for its direct detection if it forms part of our 
galactic halo. 
\end{abstract}
\end{titlepage}
\setcounter{footnote}{0}
\setcounter{page}{1}
\setcounter{section}{0}
\setcounter{subsection}{0}
\setcounter{subsubsection}{0}

\section{Introduction}

Supersymmetric models with conserved $R$-parity have in the lightest
supersymmetric particle (LSP) a natural candidate for dark matter. 
This very appealing feature has motivated much work (see 
\cite{prep,gen} for review and references) and many increasingly 
sophisticated studies on the field. Most of these analyses have 
concentrated in the minimal supersymmetric standard model (MSSM) with 
or without theoretical constraints on its wide parameter space. It is
important, however, to keep an open mind to the possibility that this 
simplest supersymmetric generalization of the standard model may not
be the one realized in nature. One should be careful not to identify 
the predictions of the MSSM with those generic of low-energy 
supersymmetry. In this respect it is healthy to explore (well 
motivated) extensions of the MSSM in search of phenomenological (or
cosmological) consequences that are different from those expected in 
the minimal model. 

Perhaps among the best motivated extended models are those that
include an additional $U(1)$ factor in the gauge group, broken 
radiatively at some scale below ${\cal O}(1)\ TeV$ by the VEV of a 
standard model singlet $S$. This type of models can be generically 
expected to arise as the low-energy limit of some string models 
\cite{cl} and have a number of interesting consequences both at the
phenomenological (e.g. for $Z'$ and Higgs physics) and theoretical 
level (e.g. they can accommodate a natural solution to the 
$\mu$-problem). For a detailed study of this kind of scenarios we 
refer the reader to refs.~\cite{cl,cdeel}. In this letter we would 
like to consider the lightest neutralino in this type of models as
a possible good dark matter candidate. We do not attempt a complete 
exploration of the parameter space of these models (even wider than 
that of the MSSM) but rather focus on a particular region in which 
the LSP has properties completely different from those that could be 
expected in the MSSM. As is described in the next section, in the 
region we study, the LSP is a neutralino mainly composed of the
fermionic superpartner of the singlet used to break the extra $U(1)$. 
The relic density of such particle, computed in section~3, turns out 
to be of the right order of magnitude for an interesting dark matter 
candidate, and the prospects for its laboratory detection are 
estimated in section~4. Finally, in section~5 we present some 
conclusions.

\section{The Dark Matter Candidate}

The symmetry breaking sector in these models includes a chiral 
multiplet $S$, singlet under the standard model gauge group but with 
a $U(1)'$ charge $Q_S$. This field couples in the superpotential
\be
\label{super}
W = h S H_1 H_2 \;\;,
\ee
to the usual Higgs doublets $H_1$, $H_2$ [with $U(1)'$ charges $Q_1$ 
and $Q_2$ respectively; gauge invariance requires $Q_1+Q_2+Q_S=0$] and
the $\mu$ parameter is dynamically generated by the VEV of $S$ as 
$\mu_s= h \langle S \rangle =h s/\sqrt{2}$.

The masses of the neutral gauge bosons $Z$ and $Z'$ are
\be
M_{Z-Z'}^2=\left[\begin{array}{cc}
\frac{1}{4}G^2(v_1^2+v_2^2)& \frac{1}{2}g_1'G(Q_1v_1^2-Q_2v_2^2)\\
\frac{1}{2}g_1'G(Q_1v_1^2-Q_2v_2^2) & {g_1'}^2(Q_1^2v_1^2+Q_2^2v_2^2+Q_S^2s^2)
\end{array}\right],
\ee
where $g$, $g'$, $g_1'$ are the gauge couplings of $SU(2)_L$, $U(1)_Y$
and $U(1)'$ respectively, and $G^2=g^2+{g'}^2$. $v_{1,2}$ are the VEVs
of $H^0_{1,2}$ with $v_1^2+v_2^2\equiv v_W^2=(246\ GeV)^2$. For 
numerical work we use ${g_1'}^2=(5/3){g'}^2$. The $Z-Z'$ mixing angle 
is constrained experimentally to be less than a few times $10^{-3}$ 
(although larger values are allowed in some cases, e.g., if the $Z'$ 
has leptophobic couplings). We will assume in this paper $Q_1=Q_2$ so 
that the requirement of a small $Z-Z'$ mixing will force $\tan\beta$ 
to be close to 1. This constraint is less stringent for larger values 
of $M_{Z'}$. 

The neutralino sector has an extra $U(1)'$ bino and an additional 
higgsino $\tilde{S}$ (the {\em singlino}) besides the four MSSM 
neutralinos. The $6\times 6$ mass matrix reads (in the basis 
$\{\tilde{B'}$, $\tilde{B}$, $\tilde{W_3}$, $\tilde{H_1^0}$,
$\tilde{H_2^0}$, $\tilde{S}\}$):
\begin{eqnarray}
\label{neutralinos}
M_{\tilde{\chi}^0}=\left(\begin{array}{c c c c c c} 
M'_1 & 0 & 0 &
 g'_1Q_1v_1 &
 g'_1Q_2v_2 &
 g'_1Q_Ss \vspace{0.1cm}\\
0 & M_1 & 0 &
-{\displaystyle\frac{1}{2}}g'v_1 &
{\displaystyle\frac{1}{2}}g'v_2 & 0\vspace{0.1cm}\\
0 & 0 & M_2 &
{\displaystyle\frac{1}{2}}gv_1 &
-{\displaystyle\frac{1}{2}}gv_2 & 0\vspace{0.1cm}\\
g'_1Q_1v_1 &
-{\displaystyle\frac{1}{2}}g'v_1 &
{\displaystyle\frac{1}{2}}gv_1 & 0 &
-\mu_s &
-\mu_s{\displaystyle\frac{v_2}{s}}\vspace{0.1cm}\\
 g'_1Q_2v_2 &
{\displaystyle\frac{1}{2}}g'v_2 &
-{\displaystyle\frac{1}{2}}gv_2 &
-\mu_s & 0 &
-\mu_s{\displaystyle\frac{v_1}{s}}\vspace{0.1cm}\\
 g'_1Q_S s & 0 & 0 &
-\mu_s{\displaystyle\frac{v_2}{s}}&
-\mu_s{\displaystyle\frac{v_1}{s}}& 0
\end{array}\right), 
\end{eqnarray}
where $M'_1$, $M_1$ and $M_2$ are the gaugino masses associated with 
$U(1)'$, $U(1)_Y$ and $SU(2)_L$ respectively. If we assume unification
of the gaugino masses at the gauge unification scale, $M'_1, M_1, M_2$
are in the proportion ${g'}_1^2 k'_1: \frac{3}{5} {g'}^2:g^2$ where 
$k'_1$ is a normalization constant. With this assumption, the
neutralino  mass matrix depends (for fixed charges) on two unknown mass
parameters, $M_1'$ and $M_{Z'}$ (through $\mu_s$) and the
dimensionless $\tan\beta$. In fig.~1 we give a contour plot of the 
lightest neutralino mass $m_\chi$ in the plane $M_1'-M_{Z'}$ for 
$\tan\beta=1$ and $Q_1=Q_2=1$. This mass is zero along a line in the
half-plane $M_1'>0$. For the study of the region around this line 
(where $\chi$ will be quite light) it is useful to define the 
normalized neutralino state $\tilde{N}$:
\bea
\tilde{N} & = & \frac{1}{N} \left[ \frac{h}{\sqrt{2}g'_1} 
r^2 \sin 2 \beta \tilde{B'}+  (\overline{Q}_H r^2-Q_S) \tilde{S} \right. 
\nonumber \\
& + & \left. r_1 (Q_S-\overline{Q}_{-}r^2) \tilde{H}_1^0 + r_2
(Q_S+\overline{Q}_{-}r^2)
\tilde{H}_2^0 
 \begin{array}{c}
\\ 
\\
\end{array}\right],
\label{N}
\eea
where $\overline{Q}_{-} = Q_1 c_{\beta}^2-Q_2 s_{\beta}^2$ (when the 
$Z-Z'$ mixing angle is small due to cancellations, $\overline{Q}_{-} 
\simeq 0$); $\overline{Q}_H = Q_1 c_{\beta}^2+Q_2 s_{\beta}^2$; 
$r_{1,2} = v_{1,2}/s$ and $r^2=r_1^2+r_2^2$. This state gives a 
precise analytic description of the $\chi$ composition near the 
$m_\chi=0$ line. This is also shown in fig.~1 where the thick solid 
line delimits the area in which the lightest neutralino has a 
$\tilde{N}$ purity $P\equiv\tilde{\chi}\cdot \tilde{N} \geq 0.99$. In 
this region $M_{Z'}$ is typically large ($M_{Z'} > 300$ GeV) so that 
$r$ is expected to be small. To first order in $r$ we can approximate:
\be
\label{pure}
\tilde{N} \simeq \frac{1}{\sqrt{1+r^2}} \left[ r(\cos \beta \tilde{H_1^0}
+\sin \beta \tilde{H_2^0}) - \tilde{S} \right] \;\;.
\ee
%
\begin{figure}[hbt]
\centerline{
\psfig{figure=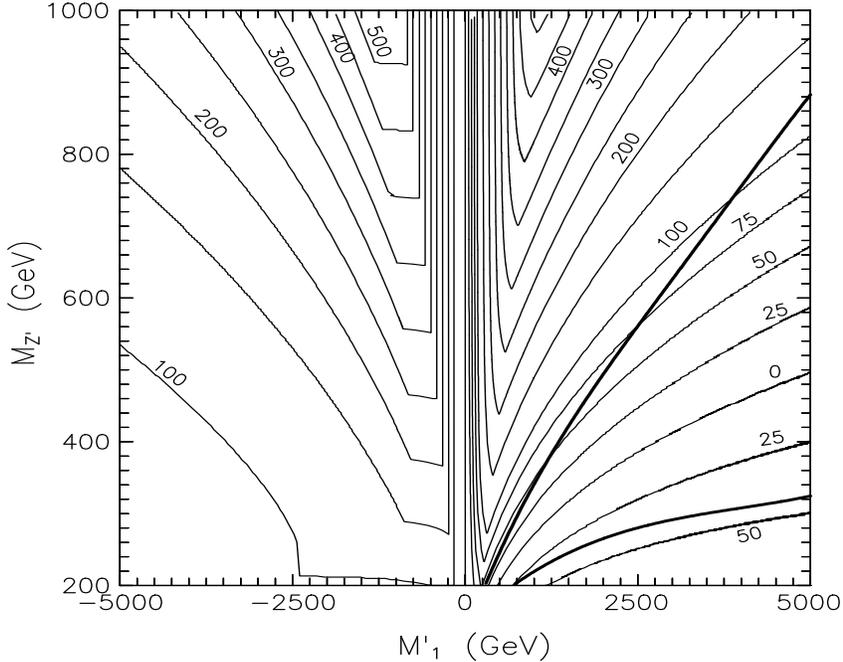,height=10cm,width=12cm,bbllx=2cm,bblly=2cm,bburx=15cm,bbury=15cm}}
\caption{Contour plot of the lightest neutralino mass ($m_\chi/GeV$)
in the plane $M_1'-M_{Z'}$ with $Q_1=Q_2=1$ and $\tan\beta=1$. In the 
region delimited by the thick solid lines, the light neutralino has 
the composition (\ref{N}) with a purity $\geq 0.99$.}
\end{figure}
%
If we write
\be
\chi=N_{1'0}\tilde{B}'+N_{30}\tilde{H}_1^0+
N_{40}\tilde{H}_2^0+N_{50}\tilde{S},
\label{compo}
\ee
we find numerically $N_{1'0}\sim 0.1-0.2$, 
$N_{30}\simeq N_{40}\sim 0.2-0.3$ and $|N_{50}|\sim 0.9-0.95$ in the 
region of interest. This is our dark matter candidate and it is 
basically singlino dominated, with a small higgsino doublet component 
and an even smaller $\tilde{B}'$ part. This result is independent of 
the assumption of unification of the gaugino masses and holds as long 
as they are large enough. More precisely, we would have obtained the 
same eq.~(\ref{N}) provided $M_1, M_2\gg M_Z$ at the electroweak scale.

The rest of neutralinos (and charginos) have masses controlled by the 
large gaugino masses and $\mu_s$ and are thus much heavier than $\chi$.  
In these models the typical soft mass scale is of order $M_{Z'}$ so 
that the spectrum of superpartners is expected to have masses of 
similar magnitude (a possible exception could be the lightest stop 
because of a sizeable $\tilde{t}_R-\tilde{t}_L$ mixing). The fact that
electroweak symmetry occurs at a lower scale (and thus $M_Z\ll 
M_{Z'}$) is a result of accidental cancellations among soft masses
(see \cite{cdeel}). The lightest scalar Higgs boson, $h^0$, remains 
also at the electroweak scale (roughly given by $M_Z$) while the rest 
of Higgs states (two more scalars $H_2^0, H_3^0$, one pseudoscalar 
$A^0$ and a charged pair $H^\pm$) have heavy masses comparable to 
$M_{Z'}$.

When $\chi$ is so light that the decay $Z \rightarrow \chi \chi$ is 
kinematically allowed it gives an extra contribution to the invisible 
$Z$ width. The LEP constraint $\delta \Gamma_{inv} < 4$ MeV 
\cite{invi} is however easily satisfied; the coupling of $\chi$ to 
the $Z$ boson is proportional to $r^2 \cos 2 \beta$ and this is small,
first, because $\tan \beta$ is close to 1 as required to suppress the
$Z-Z'$ mixing, and second, because $r$ is also small.
 
In the region of parameters described above, $\chi$ is the LSP and, 
with the assumption of conserved $R$-parity, becomes a possible 
candidate for cold dark matter.

\section{Relic Abundance}

The present relic abundance of $\chi$'s ($\Omega_\chi h^2$) is 
determined by their annihilation cross section at freeze-out, which, 
in the non-relativistic expansion applicable in this case, we write as
\be 
\sigma_{ann} v \simeq a + \frac{b}{6}v^2 \;\; ,
\label{sigma}
\ee
where $v$ is the relative velocity of $\chi$'s in the c.m. frame. The 
freeze-out temperature $T_F$ (and thus $v$) can be iteratively 
computed from the condition (obtained from the equality of 
annihilation and expansion rates):
\be
\frac{m_\chi}{T_F}\equiv x_F=\ln\frac{0.1 M_{Pl}\langle\sigma_{ann}v\rangle
m_\chi}{\sqrt{g_*x_F}}\;\;,
\ee
where $M_{Pl}=1.22\times 10^{19}\ GeV$ is the Planck mass, $g_*$ the 
number of relativistic degrees of freedom at $T_F$ ($\sqrt{g_*}\sim 
8-9$) and
\be
\langle\sigma_{ann}v\rangle=a+\left(b-\frac{3}{2}a \right)\frac{1}{x_F}\;\;,
\ee
is the thermally averaged cross section \cite{thermav}. Typically 
$x_F\simeq 20$ and $v\simeq 1/3$. For the range of masses we 
consider, $5-10\simlt m_\chi/GeV\simlt 70$, $T_F$ is above the QCD 
quark-hadron phase transition and below the electroweak phase 
transition. The neutralino relic density is then
\be
\Omega_\chi h^2 \equiv \frac{\rho_\chi}{\rho_c/h^2}= 
\frac{8.77\times 10^{-11} x_F{\rm GeV}^{-2}}{\sqrt{g_*}
\left[a+\frac{1}{2}(b-\frac{3}{2}a)\frac{1}{x_F}\right]}
\;\; , \label{omega}
\ee
where $\rho_c$ is the critical density and $h$ is the Hubble constant 
in units of $100\ km\ Mpc^{-1}\ sec^{-1}$.

$\chi$'s in that mass range can annihilate only into standard model 
fermion-antifermion pairs\footnote{Annihilation into gluons or photon
pairs is a one-loop process and can be ignored.} (top quarks 
excluded). The relevant processes are mediated by $Z$, $Z'$ and 
neutral Higgs bosons in the s-channel or by sfermions in the
t-channel. The cross section formulae for the MSSM \cite{drno} can 
be easily generalized to our model. For the particle spectrum and 
neutralino composition described previously, the cross section is 
dominated by $Z'$ exchange with ($\sigma v=\sum_f[a_f+(b_f/6)v^2]$):
\be
a_f =\frac{2c_f}{\pi} \beta_f  {g'}_1^4 \frac{m_f^2}{M_{Z'}^4}
[{Q}_f^A Q_\chi]^2\;\;,
\label{azp}
\ee
and 
\be
b_f=b_{Z'Z'}^{(f)}+\left(
-\frac{3}{2}+\frac{3}{4}\frac{\xi_f}{1-\xi_f}\right)a_f,
\ee
with
\be
b_{Z'Z'}^{(f)}=\frac{2c_f}{\pi}\beta_f\left(
\frac{{g_1'}^2Q_\chi
m_\chi}{4m_\chi^2-M_{Z'}^2}\right)^2
\left[(Q_f^V)^2(2+\xi_f)+2(Q_f^A)^2(1-\xi_f)\right]\;\; .
\ee
In this formulas, $c_f=1(3)$ for leptons (quarks) in the final state,
$\xi_f=m_f^2/m_\chi^2$, $\beta_f=\sqrt{1-\xi_f}$ and 
$Q_\chi=Q_1N_{30}^2+Q_2N_{40}^2+Q_SN_{50}^2$ for a $\chi$ with 
composition as given by eq.~(\ref{compo}). The axial and vector
$U(1)'$ charges of the final state fermions are
\be
Q_f^A=\frac{1}{2}\left[Q'(f_L)-Q'(f_R)\right]\ ,\;\;\;\;\;
Q_f^V=\frac{1}{2}\left[Q'(f_L)+Q'(f_R)\right]\ .\vspace{0.1cm}
\ee
The dependence of $\sigma_{ann}v$ on these charges introduces some 
model dependence in the results. However, annihilation into $b\bar{b}$
usually dominates so that $\Omega_\chi h^2$ depends basically on 
$Q_b^A$, which is equal to $-Q_1$ if the bottom mass is generated by 
$\langle H_1\rangle$. 

Other subdominant annihilation channels are discussed below:

$\bullet$ The amplitude for $Z^0$-mediated annihilation, which is 
usually dominant for higgsino LSP's, is proportional to 
$(N_{30}^2-N_{40}^2)$, and therefore is doubly suppressed in our case: 
$\tan\beta\simeq 1$ implies $N_{30}\simeq N_{40}$ and,
furthermore, $N_{30}^2$ and $N_{40}^2$ are very small.

$\bullet$ The t-channel contribution from sfermions, which are 
expected to have masses comparable to
$M_{Z'}$, is suppressed by extra powers of small Yukawa couplings and/or
the smallness of $N_{1'0}$, $N_{30}$, $N_{40}$. 

$\bullet$ The amplitude for $\chi\chi\rightarrow
A^0\rightarrow f \bar{f}$ has a gauge part suppressed by the smallness
of $N_{1'0}N_{30}$, $N_{1'0}N_{40}$. There is also a new part 
proportional to $hN_{50}(N_{30}\cos\beta+N_{40}\sin\beta)$ due to the 
new Yukawa coupling in the superpotential (\ref{super}) and, the 
sizeable value expected for $h$ ($h\simeq 0.7$ from renormalization 
group analyses \cite{cdeel}), can in principle compensate for the 
smallness of $N_{30}$, $N_{40}$. However, assuming $m_{A}\sim M_{Z'}$,
the $A^0$ contribution is suppressed with respect to the $Z'$ 
contribution by an extra factor $m_\chi/v_W$. 

$\bullet$ The same suppression factor appears for 
$\chi\chi\rightarrow H_2^0\rightarrow f \bar{f}$.
In addition, the $\chi\chi H_2^0$ coupling goes like $\cos 2\beta$ 
in the limit $m_A\simeq m_{H^0} \gg M_Z$ and thus is unimportant for 
$\tan\beta\simeq 1$. 

$\bullet$ In the limit $M_Z'\gg M_Z$, the third scalar, $H_3^0$, 
is singlet dominated and does not mediate the annihilation into fermions.

$\bullet$ More important can be the $h^0$ mediated annihilation; the
suppression factor $m_\chi/v_W$ can be compensated by the fact that 
$m_{h^0}$ cannot be as heavy as $M_{Z'}$ and, in addition, there can
be a resonant enhancement of the cross section for $2m_\chi\simeq 
m_h$. Nevertheless, we have checked numerically that this channel
can also be neglected. First, the CP odd nature of the $\chi\chi$ 
s-wave initial state forces $h^0$ to contribute to $\sigma_{ann}v$ 
only to order $v^2$. Second, in these models the mass of $h^0$ 
receives extra contributions and is given by
\be
m_{h^0}^2=\left[\frac{1}{4}G^2\cos^2 
2\beta+h^2\sin^22\beta+{g_1'}^2\overline{Q}_H^2\right]v_W^2+
\Delta_{rad} m_{h^0}^2,\vspace{0.1cm}
\ee
where the last piece, coming from loop corrections can also be sizeable.
One sees that, even in the case $\tan\beta=1$, the tree-level mass 
can easily be as large as $170\ GeV$, and $2 m_\chi$ is never close 
to the pole for light $m_\chi$.

$\bullet$ Finally, as we showed above, the rest of neutralinos and 
the charginos are much heavier than $\chi$ and ``co-annihilation'' 
effects (such as $\chi\chi'\rightarrow f\bar{f}$ or 
$\chi\chi^\pm\rightarrow f\bar{f}'$) play no role.

Figure~2 shows the relic density of $\chi$'s vs. its mass for 
different values of $M_{Z'}$ from 300 GeV (lower curve) to 800 GeV (upper).
\begin{figure}[hbt]
\centerline{
\psfig{figure=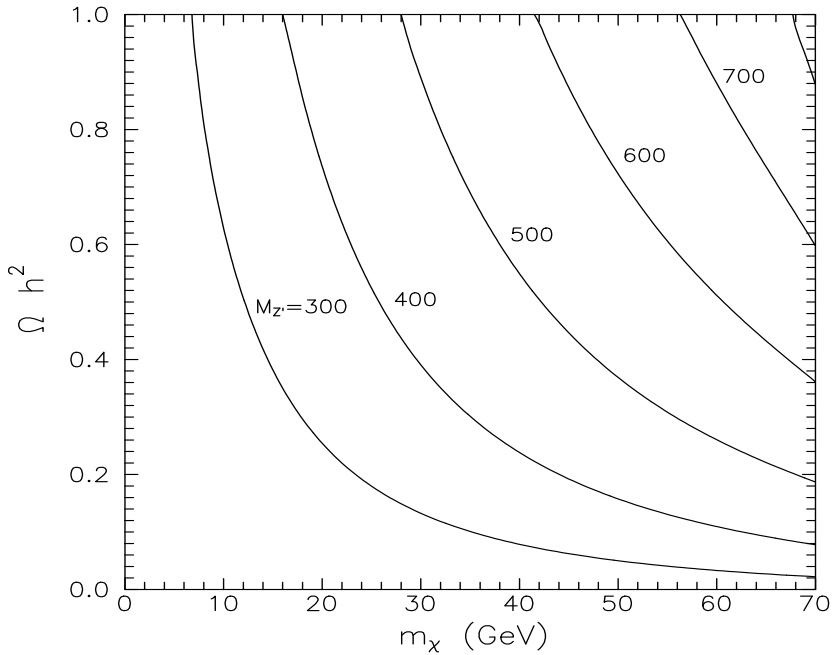,height=10cm,width=12cm,bbllx=2cm,bblly=2cm,bburx=15cm,bbury=15cm}}
\caption{Relic abundance of $\chi$ as a function of its mass (in GeV) 
for $Q_1=Q_2=1$ and different values of $M_{Z'}$ from 300 GeV 
(lower curve) to 800 GeV (upper).}
\end{figure}
%
For concreteness, we have fixed $N_{30}=N_{40}=0.25$ and 
$N_{50}=-0.9$ in eq.~(\ref{N}) and included Higgs subdominant 
annihilation channels to compute the abundance. We use $Q_1=Q_2=1$ 
and fix $m_h=170\ GeV$, and $m_A=M_{Z'}$. In the range of $m_\chi$ 
shown, $Z'$ gives the dominant annihilation channel, as explained 
above, so that larger $M_{Z'}$ reduces the annihilation rate making 
the relic abundance grow. For $m_\chi\simgt 70\ GeV$, $h^0$-exchange 
starts to be important and reduces significantly the relic density. 
Smaller values of the $U(1)'$ charges would also tend to increase the
abundance but, on the other hand, $m_h$ is lighter in that case and will
become important for lighter values of $m_\chi$.

In the figure, we show only the region $\Omega_\chi h^2 <1$, 
conservative limit which follows from a lower bound on the age of 
the Universe of $\sim 10\ Gyr$. Taking the uncertainty in the Hubble 
constant to be $0.4\simlt h \simlt 1$, the cosmologically interesting 
range is $0.01 \simlt \Omega_\chi h^2\simlt 0.5$ if $\chi$'s form a 
significant fraction of the total dark matter in the Universe.
The upper bound $\Omega_\chi h^2<1$ sets a lower limit on the value of 
$m_{\chi}$ for which $\chi$ is a good CDM candidate, given fixed 
values of the charges and $M_{Z'}$. This lower bound increases with 
$M_{Z'}$ and can be very small for low $M_{Z'}$ without conflicting 
with experimental bounds from $Z^0$ decays.

A non-negligible $\chi\chi\rightarrow Z'\rightarrow f\bar{f}$ rate 
usually implies a sizeable $Z'$ production cross section at hadron 
colliders and one should check that the interesting region for 
$\Omega_\chi h^2$ is not in conflict with the non-observation of 
$Z'$ events (basically $p\bar{p}\rightarrow Z'\rightarrow e^+e^-$) at 
the Tevatron CDF and D0 experiments. The lower limits on $M_{Z'}$ 
obtained by these experiments are model-dependent (all the $U(1)'$ 
charges and masses of the possible decay products of $Z'$ enter the 
computation) and we do not attempt such detailed analysis. We remark, 
however, that the supersymmetric decays of the $Z'$ (in particular 
the invisible decay $Z'\rightarrow \chi\chi$ can be very important)
reduce significantly the $Z'\rightarrow e^+e^-$ branching ratio, so 
that the usual limits on $M_{Z'}$ are relaxed \cite{susyzp} and can be
easily evaded in our model.

It is illustrative to compare our dark matter candidate with similar 
light higgsino candidates proposed in other supersymmetric models. In 
the MSSM one such light neutralino, with composition $\chi\simeq 
\sin\beta \tilde{H}_1^0+ \cos\beta\tilde{H}_2^0$, was studied in 
\cite{kawe} motivated by the SUSY interpretation of the CDF 
$ee\gamma\gamma+\slashchar{E}_T$ event. The $\chi\chi Z^0$ coupling, 
proportional to $\cos2\beta$, is reduced for $\tan\beta\simeq 1$. 
This suppresses the contribution to the invisible $Z^0$ width and the 
$\chi\chi\rightarrow Z\rightarrow f \bar{f}$ cross-section, producing 
relic abundances in the interesting range.

In the NMSSM, the minimal model extended by an extra singlet but no 
additional $U(1)$, another light higgsino dark matter candidate was 
studied in \cite{oletal} (for more general analyses of NMSSM 
neutralino dark matter see \cite{nmssm}). In this case $\chi\simeq 
(\cos\beta \tilde{H}_1^0+ \sin\beta\tilde{H}_2^0+\alpha \tilde{S})/N$ 
with all three components of similar magnitude. The contribution to 
$\Gamma_Z^{inv}$, also proportional to $\cos^22\beta$ in this case, 
is further reduced by the non-negligible $\tilde{S}$ component. If 
$\chi\simeq\tilde{S}$, the relic abundance would be too large 
because $\tilde{S}$'s do not annihilate efficiently.

Our model is similar to the previous case but with $\tilde{S}$ 
dominant in the $\chi$ composition. Now, however, $\chi$'s annihilate 
through $\chi\chi\rightarrow Z'\rightarrow f\bar{f}$. The rate is not
too large due to the smallness of $g_1'Q_S$ and the large mass of the 
$Z'$ boson, resulting in a relic abundance of the right order of 
magnitude. 

Finally, some comments from the model building point of view are in 
order. The region of parameter space we have examined corresponds to 
$M_1'\gg M_{Z'}$ and $M_1,M_2\gg M_Z$. Such hierarchy of masses 
cannot be easily accommodated in models with universal boundary 
conditions at a high energy scale (a GUT scale or the string scale)
and would rather point to models in which soft breaking is dominated
by gaugino masses. Models of this kind (in the context of the MSSM) 
have been considered in the past \cite{gaugino}. A renormalization 
group analysis of the evolution of parameters from low-energy to the 
string scale, with particular attention to symmetry breaking 
constraints, would be required.

\section{Detection Rates}

If $\chi$'s form the bulk of the dark matter in our galactic halo 
(with density $\rho_0\sim 0.3\ GeV/cm^3$ measured by its gravitational
effects), we would like to estimate the prospects for its detection. 
We focus on direct detection experiments, which are the best suited 
for small mass WIMPs. In these experiments one hopes to detect 
calorimetrically nuclear recoils in specialized materials after 
elastic scattering with the flux of $\chi$'s. In the limit $N_{50}=1$,
the (spin-dependent) $\chi$-nucleus scattering proceeds by $Z'$ 
exchange and the rate of events per day and kg of material is \cite{prep}
\be\label{rate}
R=\frac{\sigma_{sd}\rho_0}{m_\chi m_N}(\xi v)_{sd}\left[\frac{7.3\times
10^{55}}{kg\cdot day }\right],
\ee
where $m_N$ is the nucleus mass, $(\xi v)_{sd}$ takes into account 
the nuclear form factor suppression and velocity distribution of 
$\chi$'s, and \cite{prep}
\be
\sigma_{sd}=\frac{16}{\pi}m_r^2\frac{J(J+1)}{J^2}\left[
\sum_{q=u,d,s}d_q(\Delta q^{(p)}\langle S_p\rangle+
\Delta q^{(n)}\langle S_n\rangle)
\right]^2.
\ee
In this formula, $J$ is the spin of the nucleus, 
$m_r=m_Nm_\chi/(m_N+m_\chi)$ is the reduced mass of the 
$\chi$-nucleon system, $\Delta q^{(p)}$ ($\Delta q^{(n)}$) is the 
quark spin content of the proton (neutron), $\langle S_{p,n}\rangle$ 
is the expectation value of the spin content of the proton (or 
neutron) group in the nucleus and finally
\be
d_q={g_1'}^2\frac{Q_q^AQ_\chi}{M_{Z'}^2},
\ee
gives the effective neutralino-quark axial coupling.

For non-zero $N_{30}$, $N_{40}$ there is also a scalar interaction 
mediated by $h^0$ exchange. The corresponding rate is of the form 
(\ref{rate}) with a different form factor $(\xi v)_{sc}$ and 
$\sigma_{sd}$ replaced by
\be
\sigma_{sc}=\frac{4}{\pi}m_r^2\left[
Zf_p+(A-Z)f_n
\right]^2.
\ee
$Z$ and $A$ are the nuclear charge and atomic number respectively. 
$f_p$ and $f_n$, the effective scalar couplings of $\chi$ to protons 
and neutrons, are proportional to
\be
f_h^{(p,n)}=\frac{hm_{p,n}}{\sqrt{2}v_W}N_{50}(N_{30}\sin\beta+
N_{40}\cos\beta)\frac{1}{m_h^2},
\ee
and $m_{p,n}$ are the proton and neutron masses, so that $f_p\simeq 
f_n$. For heavy nuclei, the $A^2$ factor can compensate for the 
smallness of $N_{30}$, $N_{40}$, and $\sigma_{sc}$ eventually 
dominates over the spin-dependent rate.

We have computed the total rate (spin-dependent plus scalar) for two 
typical materials used in dark matter detectors, $^{73}$Ge and 
$^{29}$Si, both with $J\neq 0$. The results, as a function of 
$m_\chi$, are presented in fig.~3 for the same choice of parameters 
used in fig.~2. In each set of curves, the upper one corresponds to 
$M_{Z'}=300\ GeV$ and the lowest to $M_{Z'}=800\ GeV$. For 
$M_{Z'}>400\ GeV$ the curves are nearly indistiguishable:
the spin-dependent rate is negligible and the scalar rate (independent
of $M_{Z'}$) gives all the effect. The small difference between the
two limiting curves in each set shows that the scalar interaction 
provides the dominant effect.
\begin{figure}[hbt]
\centerline{
\psfig{figure=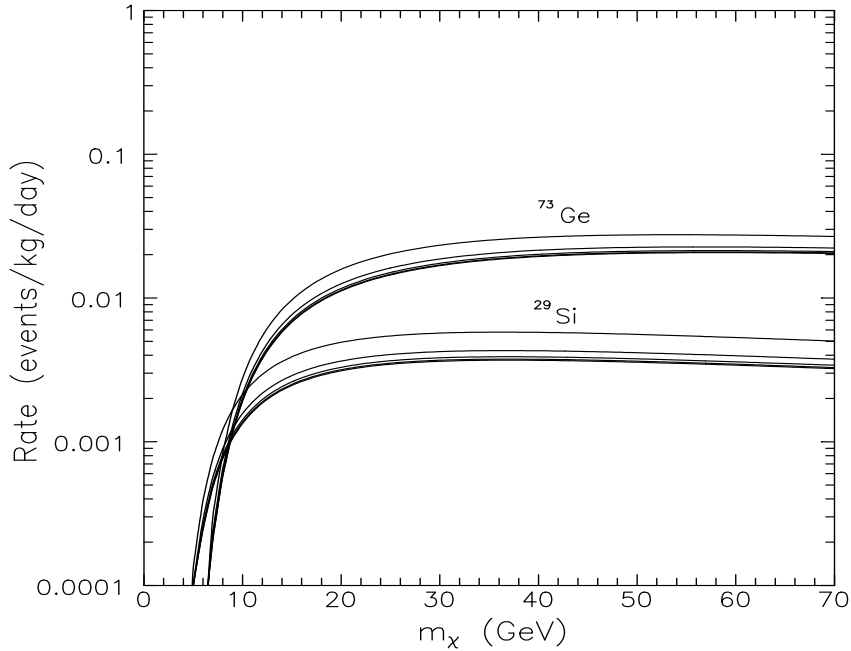,height=10cm,width=12cm,bbllx=2cm,bblly=2cm,bburx=15cm,bbury=15cm}}
\caption{Detection rates for $^{73}$Ge and $^{29}$Si detectors.}
\end{figure}
The rates obtained are below the current experimental sensitivity 
\cite{prep}. For instance the CDMS experiment, using germanium, is
planning on obtaining a sensitivity of $\sim 10^{-1}$ event/kg/day 
after a year of exposure. However, the next generation of experiments 
is expected to achieve sensitivities $\sim 0.01$ event/kg/day and 
therefore, we conclude that $\chi$'s could well be on the reach of 
near future experiments. 

\section{Conclusions}

In supersymmetric models with an extra $U(1)$ (broken at the TeV scale
by the VEV of an standard model singlet $S$) the LSP neutralino can be
singlino dominated. We have showed that the thermal relic density of 
this fermion, stable if $R$-parity is conserved, can be of the right 
order of magnitude to be a good cold dark matter candidate.
We have also estimated direct detection rates at typical cryogenic 
devices in search of halo dark matter and found that they are 
typically small but may be reachable by the next generation of 
experiments.

\section*{Acknowledgements}

We thank L. Everett, M. Hindmarsh, R. Jeannerot and P. Langacker 
for help and useful discussions. B. de C. thanks the Dept. of Physics 
and Astronomy of the University of Pennsylvania for their warm 
hospitality and financial support during the realization of this work.

\end{document}